\documentclass[12pt]{article}
\usepackage{epsfig}
\usepackage{psfig}
\usepackage{graphicx}
\usepackage{amsmath}
\usepackage{amsfonts}
\usepackage{amssymb}

\begin{document}

\date{}
\title{\textbf{Dynamical Casimir effect with Dirichlet and Neumann boundary conditions}}
\author{D. T. Alves$^{1,2}$, C. Farina$^{3}$ and P. A. Maia Neto$^{3}$\\{\small \textit{$^{1}$Centro Brasileiro de Pesquisas F{\'\i}sicas, Rua Dr.
Xavier Sigaud 150, Rio de Janeiro RJ, 22290-180, Brasil }}\\{\small \textit{$^{2}$ Departamento de
F{\'\i}sica, Universidade Federal do Par\'{a}, Bel\'{e}m  PA, 66075-000, Brasil }}\\{\small \textit{$^{3}$Instituto de F{\'\i}sica, UFRJ, Caixa Postal
68528, Rio de Janeiro RJ, 21945-970, Brasil}}}
\date{\today}
\maketitle
\begin{abstract}
We derive the radiation pressure force on a non-relativistic moving plate in 1+1 dimensions.
We assume that a massless scalar field satisfies either Dirichlet or  Neumann 
boundary conditions (BC) at the instantaneous position of the plate. 
We show that when the state of the field is invariant under time translations, the results derived for
Dirichlet and Neumann BC are equal. We discuss the force for a thermal field state as an example 
for this case. On the other hand, 
a  coherent state introduces a phase reference, 
and the two types of BC lead to different results.
\end{abstract}

\section{Introduction}

An intriguing feature of the static Casimir effect is the dependence of the force on the 
type of boundary conditions imposed on the field. For two parallel infinitely permeable plates 
the force turns out to be attractive, and in fact identical to the original Casimir result for 
a pair of perfectly conducting plates. 
On the other hand, the Casimir force between a perfectly conducting plate and a permeable one 
is, surprisingly, repulsive~\cite{Boyer}. 
Due to this peculiarity, permeable plates have been considered recently in the literature in the context of Casimir effect as well as Cavity QED  \cite{permeableplatesrecently}.

An analogous situation takes place for a scalar field in 1+1 dimensions:  
taking Neumann  conditions at the boundaries results in the same attractive force obtained with 
Dirichlet BC, whereas in the mixed case the force is 
repulsive~\cite{Boyer1+1}. 
For  reviews on the Casimir effect see \cite{reviewstat} and references therein.

In the case of moving boundaries~\cite{Moore}, the force exerted by vacuum fluctuations 
usually contains a dissipative component~\cite{dissipation}. 
The amount of mechanical energy dissipated  
is converted into pairs of real particles~\cite{emission}.
These radiation reaction forces on moving
bodies may appear even in the case of only one moving wall~\cite{FDavies}--\cite{PA94}. 
Connections between the dynamical Casimir effect and several interesting phenomena such as 
the Unruh-Davies effect~\cite{Unruh-Davies},
black hole physics~\cite{black_hole}, sonoluminescence~\cite{sono}, mass corrections~\cite{mass}
and quantum decoherence~\cite{PAprl} have been discussed 
in the literature~\cite{review}.

In this paper, we extend the discussion of the role played by different BC to the dynamical effect. 
We compute explicitly the radiation pressure force for both Dirichlet and Neumann BC, and discuss
the class of field states for which the force is the same for these two types of BC. 
We consider a single moving boundary in the nonrelativistic approximation, and a 
massless scalar field in 1+1 dimensions in a general quantum state. This paper is organized as follows:
Section 2  presents the results for  
Dirichlet BC. In Section 3, we consider Neumann BC and 
show that 
these two BC lead to the same force 
when the field state is invariant under time translations. 
As an example of such a state, we discuss
the case of a thermal field in Section 4. 
In Section 5, we consider a coherent state, which is
an example of field state that does not satisfy this symmetry.  
Section 6 is devoted to the conclusions and final remarks.

\section{ Dirichlet BC }

In this section, we  assume that the field vanishes 
at the instantaneous position of the plate, when measured in the Lorentz frame $S'$ 
which is co-moving at a given time:
$\phi ^{\prime }(x^{\prime},t^{\prime })\vert_{plate}=0, 
$
where the prime quantities refer to $S'.$ 
In terms of the laboratory coordinates, this BC is equivalent to~\cite{Moore}
\begin{equation}
\phi(\delta q(t),t)=0. \label{Dirichlet}
\end{equation}
We solve eq.~(\ref{Dirichlet}) in the 
long wavelength approximation, and assume the effect of the 
motion  to be a small perturbation. We follow Ford and Vilenkin~\cite{Ford}, and define 
\begin{equation}\label{FV}
\phi (x,t)=\phi _{0}(x,t)+\delta \phi (x,t),
\end{equation}
where $\phi _{0}$ corresponds to the solution with a static plate at the origin and $\delta\phi$ is a small perturbation which takes into account the effect of the motion of the plate. Hence the unperturbed field $\phi_0$ satisfies the wave equation $\square\phi_0(x,t)=0$ and the BC  $\phi_0(0,t)=0.$ Its normal mode expansion 
in the half-space $x>0$ 
is\footnote{We are using $c=1$ along the text.}
\begin{equation}\label{modes}
\phi _{0}(x,t)=\int_{0}^{\infty }d\omega \sqrt{\frac{\hbar }{\pi \omega }}\,\sin (\omega x)%
\left[ a_{\omega}\,e^{-i\omega \, t}+a_{\omega}^{\dagger }\,e^{i\omega \, t}\right],
\end{equation}
with
\begin{equation}\label{commutator}
[a_{\omega},a_{\omega'} ^{\dagger }]=\delta(\omega-\omega').
\end{equation}
A similar decomposition is written in the half-space $x<0,$ with 
the bosonic operators $a_{\omega}$ and their hermitian conjugates $a_{\omega}^{\dagger }$
replaced by independent operators  $b_{\omega}$ and $b_{\omega}^{\dagger }.$

The field operator $\delta\phi$ also satisfies the wave equation $\square\delta\phi=0$ and is 
submitted to a BC which can be obtained directly from eqs.~(\ref{Dirichlet}) and (\ref{FV}) by taking the 
Taylor expansion around $x=0$ and 
neglecting terms of second order in $\delta q(t):$
\begin{equation}
\delta \phi (0,t)=-\delta q(t)\,\partial _{x}\phi _{0}(0,t).\label{D1}
\end{equation}

The net force on the moving plate may be written in terms of the suitable 
component of the energy-momentum tensor as follows:
\begin{equation}
F(t)=\left\langle T^{11}(\delta q^-(t),t)-T^{11}(\delta q^+(t),t)\right\rangle,\label{F1}
\end{equation}
with
\begin{equation}\label{BM_T}
T^{11}(x,t)=\frac{1}{2}\left[(\partial _{x}\phi)^2 (x,t)+(\partial
_{t}\phi)^2 (x,t)\right]
\end{equation}
and $\delta q^\pm=\delta q\pm\epsilon$ (with $\epsilon\rightarrow 0^+$).   
The average $\langle ... \rangle$ is taken over an arbitrary field state. For simplicity, 
however, we assume that the state in the half-space $x>0$ is defined with respect 
to the operators $a_{\omega}$ and $a_{\omega}^{\dagger }$ exactly as 
the state for the half-space $x<0$ is defined in terms of 
$b_{\omega}$ and $b_{\omega}^{\dagger }.$ In this symmetrical case,  
the net force vanishes to zero order of $\delta q$ (plate at rest at the origin). 
As an example, if we consider the field to be at thermal equilibrium at temperature
$T$ at the half-space $x>0,$ then our assumption implies that the field at the 
left-hand side of the plate is also at thermal equilibrium at the same temperature.  

For Dirichlet BC, the term 
$(\partial_{t}\phi)^2$ in (\ref{BM_T}) does not contribute to the force, and we find 
\begin{equation}
T^{11}_D(\delta q^+(t),t)=
\frac{1}{2} \left[(\partial_{x}\phi_0)^2(0^+,t)+
\biggl\{\partial_x\phi_0(0^+,t),\partial_x\delta\phi(0^+,t)\biggr\}+
{\cal O}(\delta q^2)\right],\label{T11}
\end{equation}
where $\{A,B\}$ represents the anti-commutator of two given operators $A$ and $B.$

The Fourier representation allows for an interesting physical interpretation of the dynamical
Casimir effect. Moreover, it provides a simple ultra-violet regularization of the  
force (alternatively, one may
employ the point-splitting method in the time domain~\cite{Ford}). When taking the Fourier transform 
of (\ref{F1}), we replace $T^{11}$ by the r.-h.-s. of (\ref{T11}). As discussed above, the terms independent of 
$\delta q$ from each side of the plate cancel in the symmetrical case. On the other hand,  
the linear terms  in $\delta q$ add to give
\begin{equation}
{\cal F}(\omega)=-\int\frac{d\omega^{\,\prime}}{2\pi}\,
 \langle  
\left\{\partial_x\Phi_0(0^+,\omega^{\,\prime}),
\partial_x\delta\Phi(0^+,\omega-\omega^{\,\prime})\right\}\rangle.\label{Fomega}
\end{equation}
The integration variable $\omega'$ in (\ref{Fomega}) is the frequency of the 
unperturbed field $\Phi_0;$ we avoid any  change of variable of integration, so as to 
conserve its physical interpretation. 

We now solve for the perturbed field $\delta\Phi$ in terms of  $\Phi_0.$
We take the solution that propagates outwards from the plate: $\delta\Phi(x,\omega)=e^{i\omega |x|}\delta \Phi(0,\omega),$
and from the Fourier transform of the r.-h.-s. of (\ref{D1})
we derive
\begin{equation}\label{deltaphi}
\partial_x \delta\Phi(0^+,\omega)=-i\omega\int \frac{d\omega'}{2\pi} \delta Q[\omega-\omega']
\partial_x \Phi_0[0^+,\omega'].
\end{equation}
Eq.~(\ref{deltaphi}) shows that the scattering of a given 
field Fourier component $\omega'$ generates a new frequency (or sideband) at frequency 
$\omega'+\Omega,$ where $\Omega$ is the mechanical frequency.
We replace eq.~(\ref{deltaphi}) into (\ref{Fomega}) to find
\begin{equation}
{\cal F}_D(\omega) = i \int \frac{d\omega'}{2\pi} (\omega-\omega') \int \frac {d\omega''}{2\pi}
\delta Q(\omega-\omega'-\omega'') \sigma_D(\omega',\omega''), \label{Fomega2}
\end{equation}
where we have defined the correlation function of the unperturbed field operator
\begin{equation}\label{sigmaD_def}
\sigma_D(\omega,\omega')=\langle \left\{\partial_x \Phi_0(0^+,\omega), 
\partial_x \Phi_0(0^+,\omega')\right\}\rangle.
\end{equation}
For future reference, we  derive, by
taking the Fourier transform of (\ref{modes}), the result
\begin{equation}\label{phixomega}
\partial_x\Phi_0(0^+,\omega)=\sqrt{4\pi\hbar|\omega|}
\left[\theta(\omega) a_{\omega}+
\theta(-\omega)a^{\dagger}_{-\omega}\right].
\end{equation}
From (\ref{phixomega}), we may derive the correlation function $\sigma_D(\omega,\omega')$ and hence the 
dynamical Casimir force for a variety of field states.  However, before considering some
specific examples, we consider in the next
section the case of Neumann BC.

\section{Neumann BC}

In this section we assume that space derivative of the field,
taken in the instantaneously co-moving Lorentz frame,
 vanishes 
at the plate's instantaneous position: 
\begin{equation}
\partial _{x^{\prime }}\phi ^{\prime }(x^{\prime},t^{\prime })\vert_{plate}=0.
\end{equation}
In the nonrelativistic approximation, and using 
the appropriate Lorentz transformation, this BC can be written in terms of quantities in the inertial frame of the laboratory as follows~\cite{PA94}:
\begin{equation}\label{p_cond_N}
\biggl\{\partial _{x}+\delta {\dot q}(t)\text{ }\partial _{t}\biggr\} \phi
(x,t)|_{x=\delta q(t)}=0.
 \end{equation}
As in the previous section, we write $\phi=\phi_0+\delta\phi,$ but now the unperturbed field 
is given by 
\begin{equation}\label{p_sol_estatica_N}
\phi _{0}(x,t)=\int_{0}^{\infty }d\omega \sqrt{\frac{\hbar }{\pi \omega }}\,\cos (\omega x)%
\left[ a_{\omega}\,e^{-i\omega \, t}+a^{\dagger }_{\omega}\,e^{i\omega \, t}\right] \text{,}
\end{equation}
whereas the motion induced correction satisfies, up to first order in $\delta q(t)$ and its time-derivatives,
\begin{equation}
\partial _{x}\delta \phi (0,t)=-\delta q(t)\,\partial _{x}^{2}\phi _{0}(0,t)-
\delta {\dot q}(t)\text{ }\partial _{t}\phi _{0}(0,t)\text{,}
\label{p_cond_delta}
\end{equation}
which follows from the Taylor expansion of (\ref{p_cond_N}).

The force on the plate is computed from the energy-momentum tensor as discussed in the previous section.
However, in contrast to the case of Dirichlet BC,  
here only the term $(\partial_t\phi)^2$ contribute
in the expression of the r.-h.-s. of eq.~(\ref{BM_T}), since (\ref{p_cond_N}) yields
$$
(\partial_x\phi)^2(\delta q(t),t)=\delta {\dot q}(t)^2 (\partial_t\phi)^2(\delta q(t),t)
\sim {\cal O}(\delta {\dot q}(t)^2)\; .
$$
Hence, the net force on the moving plate can be written as:
\begin{equation}\label{forcaDin1N}
F_N=-\frac{1}{2}\langle (\partial_{t}\phi)^2(\delta
q^+(t),t)-(\partial_{t}\phi)^2 (\delta q^-(t),t)\rangle.
\end{equation}
Keeping only the linear terms in $\delta q$, we obtain:
\begin{equation}\label{forcaDin2N}
F_N(t)= -\frac{1}{2}\langle  
\biggl\{\partial_t\phi_0(0^+,t),\partial_t\delta\phi(0^+,t)\biggr\}-
\big\{0^+\rightarrow 0^-\bigr\}\rangle+{\cal O}(\delta q^2).
\end{equation}
As before, the contributions coming from terms involving only the field operator $\phi_0$ cancel out, and 
the linear contributions 
from each side of the plate are equal, so that the net force on the plate can be written in the 
frequency domain as
\begin{equation}\label{forcaFourierN1}
{\cal F}_N(\omega)=\int\frac{d\omega^{\,\prime}}{2\pi}\,
 (\omega-\omega^{\,\prime})\,\omega^{\,\prime}\,\langle 
\biggl\{\Phi_0(0^+,\omega^{\,\prime}),
\delta\Phi(0^+,\omega-\omega^{\,\prime})\biggr\}\rangle\; .
\end{equation}
It is now convenient to express $\delta\Phi(0^+,\omega^{\,\prime})$ in terms of  $\partial_x\delta\Phi(0^+,\omega^{\,\prime})$, since this last quantity can be written with the help of  eq.~(\ref{p_cond_delta}) (after a Fourier transformation) in terms of the unperturbed field $\Phi_0.$
The outward solution of the wave equation given the value 
of $\partial_x\delta\Phi(0,\omega)$ is
\begin{equation}\label{ecuacion}
\delta\Phi(x,\omega)=\epsilon(x)\, \partial_x\delta\Phi(0,\omega)\,
\frac{e^{i\omega\vert x\vert}}{i\omega},
\end{equation}
where $\epsilon$ denotes the sign function. 
Substituting eq.~(\ref{ecuacion}) into eq.~(\ref{forcaFourierN1}), we obtain:
\begin{equation}\label{forcaFourierN2}
{\cal F}_N(\omega)=-i\int\frac{d\omega^{\,\prime}}{2\pi}\,
 \omega^{\,\prime}\,\langle 
\biggl\{\Phi_0(0^+,\omega^{\,\prime}),
\partial_x\delta\Phi(0,\omega-\omega^{\,\prime})\biggr\}\rangle\; .
\end{equation}
As
a last step, we take the 
 Fourier transform of eq.~(\ref{p_cond_delta}) and replace the result 
into (\ref{forcaFourierN2}):
\begin{equation}\label{forcacorrelacoes}
{\cal F}_N(\omega)=
-i\int\frac{d\omega^{\,\prime}}{2\pi}
\int\frac{d\omega^{\,\prime\prime}}{2\pi}(\omega-\omega^{\,\prime})
\omega'
\omega^{\,\prime\prime}\, 
\delta Q(\omega-\omega^{\,\prime}-\omega^{\,\prime\prime})
\sigma_N(\omega^{\,\prime},\omega^{\,\prime\prime})\; ,
\end{equation}
where we defined the correlation function of the unperturbed field operator
\begin{equation}
\sigma_N(\omega,\omega^{\,\prime})=
\langle \biggl\{\Phi_0(0^+,\omega),\Phi_0(0^+,\omega')
\biggr\}\rangle.\label{C3}
\end{equation}
in analogy with (\ref{sigmaD_def}).
The unperturbed field appearing in 
(\ref{C3}) 
is computed from (\ref{p_sol_estatica_N}): 
\begin{equation}\label{phixomega2}
\Phi_0(0,\omega)=\sqrt{\frac{4\pi\hbar}{|\omega|}}
\left[\theta(\omega) a_{\omega}+
\theta(-\omega)a^{\dagger}_{-\omega}\right].
\end{equation}
By inspection of eqs.~(\ref{phixomega}) and  (\ref{phixomega2}), we derive the following
relation between the correlation functions for Dirichlet and Neumann boundary conditions:
\begin{equation}\label{relation}
\sigma_D(\omega,\omega')=|\omega \omega'|\, \sigma_N(\omega,\omega').
\end{equation} 
This result allows us to compare the forces for Dirichlet and Neumann boundary conditions. 
When the field state is invariant under time translations,  the correlation 
function $\langle \phi_0(0,t)\phi_0(0,t') \rangle$ is a function of $t-t'$ only, and then 
in the frequency domain it satisfies $\sigma(\omega,\omega')\propto \delta(\omega+\omega').$
Hence we may replace $|\omega \omega'|=-\omega \omega'$ in (\ref{relation}), 
and from (\ref{Fomega2}) and 
(\ref{forcacorrelacoes}) it follows that Dirichlet and Neumann BC provide the same result:
\begin{equation}
{\cal F}_N(\omega)={\cal F}_D(\omega)={\cal F}(\omega)
\end{equation}
Moreover, since $\omega'=-\omega''$ in (\ref{Fomega2}) and 
(\ref{forcacorrelacoes}), 
${\cal F}(\omega)$ is proportional to $\delta Q(\omega),$
and may be written in terms of a susceptibility function ${\cal \chi}(\omega)$ as follows: 
\begin{equation}\label{chi}
{\cal F}(\omega)={\cal \chi}(\omega)  \delta Q(\omega).
\end{equation}
In the time domain, (\ref{chi}) reads
\begin{equation}
{ F}(t)=\int dt' {\tilde \chi}(t-t') \delta Q(t'),
\end{equation}
where ${\tilde \chi}(t)$ is the inverse Fourier transform of $\chi(\omega).$
Thus, the effect of the plate displacement at time $t'$ on the force at time $t$ 
depends only on $t-t',$ as expected from the assumption of time translational symmetry.  

As an example of field state obeying this assumption, we consider in the next section 
the force for a thermal field (temperature $T$). 
This example contains  
the case of the vacuum state 
as the limit $T=0.$
In Section 5, we compute the force for a coherent state (amplitude $\alpha$). This example also contains the 
vacuum state as a limiting case ($\alpha=0$), but this time we find 
different results for the two types of BC
when $|\alpha|>0.$

\section{Thermal state}

We compute in the Appendix the correlation 
function $\sigma_D$ for a thermal field:
\begin{equation}\label{sigmaD}
\sigma_D(\omega,\omega')=4\pi\hbar|\omega|
(1+2\overline{n}(\omega))\,\delta(\omega+\omega'),
\end{equation}
with  
\begin{equation}\label{nT}
\overline{n}(\omega)=[\exp(\hbar|\omega|/k_BT)-1]^{-1}
\end{equation}
representing  the average photon number at frequency $\omega$ 
($k_B$ is the Boltzmann constant). 
As expected and discussed in the previous section, $\sigma_D(\omega,\omega')$ is proportional to
$\delta(\omega+\omega'),$ 
a signature of time translational symmetry.
Hence the force is the same for Dirichlet and Neumann BC, and it is given in terms of 
a susceptibility function $\chi(\omega)$ according to (\ref{chi}).
From (\ref{Fomega2}) and (\ref{sigmaD}) we find 
\begin{equation}\label{chi2}
\chi(\omega)=2i\hbar \int \frac{d\omega'}{2\pi} |\omega'| (\omega + \omega')
\left[\frac{1}{2}+\overline{n}(\omega')
\right].
\end{equation}
Thus the susceptibility is the sum of two contributions: $\chi=\chi_{\rm vac}+\chi_T,$ where 
$\chi_T$ is  proportional to $\overline{n}(|\omega'|)$ in (\ref{chi2}),
whereas $\chi_{\rm vac}$ contains the effect of vacuum fluctuations [the term `1/2' in 
(\ref{chi2})]. 
At the zero-temperature limit $\overline{n}=0$ and then $\chi=\chi_{\rm vac}.$

Taking $\omega>0$ (but an analogous argument is valid for $\omega<0$) we can extract a finite result for $\chi_{\rm vac}(\omega)$ adopting the following regularization prescription:
\begin{equation}\label{creg}
\chi_{\rm vac}(\omega)=
i\hbar \int_{-\infty}^{+\infty }\frac{d\omega'}{2\pi}|\omega'| (\omega + \omega')
=i\hbar  \lim\limits_{\Lambda\rightarrow\infty }\left(\int_{-\Lambda-\omega}^{0} +\int_{-\omega}^{0}+\int_{0}^{\Lambda}\right)
\frac{d\omega'}{2\pi}|\omega^{\,\prime}| \,(\omega+\omega ^{\,\prime }).
\end{equation}
Since the integrand is odd under reflection around $\omega'=-\omega/2,$ 
 the first and third integrals in
 the rhs of  eq.~(\ref{creg}) cancel exactly for any value of $\Lambda.$ 
Therefore, the single relevant contribution 
is the second term, 
which corresponds to 
negative frequencies  that give rise
to sidebands with positive frequencies (on the other hand, when considering  
a negative mechanical frequency $\omega,$ the sidebands are down shifted,
and the contribution comes from the frequency interval $[0,-\omega]$).
Positive and negative frequencies correspond to annihilation and creation 
operators, hence their mixture is clearly connected to the emission of 
photons out of the vacuum state. 
From eqs.~(\ref{chi}) and (\ref{creg}), we find 
\begin{equation}\label{chivacres}
\chi_{\rm vac}(\omega )= i\hbar\omega^3/(6\pi),
\end{equation}
or, in the time domain, $F_{\rm vac}=$
{\large$\frac{\hbar}{6\pi}\frac{d^3}{dt^3}$}$\delta q(t)$, a result first obtained in Ref.~\cite{Ford}, and which corresponds to the nonrelativistic 
limit of the exact expression of Fulling and Davies~\cite{FDavies}. 

For the contribution of thermal photons, 
using that $\overline{n}(\omega')=\overline{n}(-\omega')$ we find from (\ref{chi2})
\begin{equation}\label{chiT}
\chi_T(\omega)=4i\hbar\omega \int_0^{\infty} \frac{d\omega'}{2\pi} 
\frac{\omega'}{\exp(\hbar\omega'/k_BT)-1}. 
\end{equation}
The integral in (\ref{chiT}) may be calculated in terms of 
the Riemann zeta function: 
\begin{equation}\label{susceptibilidadetermicafinal}
\chi_T(\omega)=i\,\frac{2\pi(k_BT)^2}{3\hbar}\,\omega.
\end{equation}
Adding up the results of (\ref{chivacres}) and (\ref{susceptibilidadetermicafinal}) we find
in the time domain
\begin{equation}\label{thermaldissipativeforce}
F(t)=\frac{\hbar}{6\pi} \frac{d^3}{dt^3}\delta q(t)-\frac{2\pi(k_BT)^2}{3\hbar} 
\frac{d}{dt}\delta q(t),
\end{equation}
in agreement with Ref.~\cite{JaekelReynaudPLA93}. By direct comparison 
of (\ref{chivacres}) and (\ref{susceptibilidadetermicafinal}), the viscous, thermal contribution 
in (\ref{thermaldissipativeforce}) is much larger than the vacuum force 
when the typical Fourier components of $\delta q(t)$ are 
much smaller than $k_B T/\hbar,$ which is of the order of $10^{13} {\rm s}^{-1}$
at room temperatures.  

The thermal contribution does not result from the emission of new particles only. It is
in part an effect of Doppler shifting the frequencies of the incoming thermal photons.
The counterpropagating photons are up shifted when reflected by the plate by the amount 
$\delta \omega=2\omega \delta {\dot q}(t),$ 
whereas the 
co-propagating  photons are down shifted by the same amount. By momentum conservation, in both cases the plate recoils along the direction opposite to its motion. The resulting force is proportional to the total power $P$ incident on both sides of the plate:
\begin{equation}\label{Doppler}
F_{\rm Doppler}=-2 P \delta {\dot q}(t).
\end{equation}
In the three-dimensional case, $P$ is proportional to $T^4$ (Stefan-Boltzmann law)
and so is the thermal viscous force~\cite{PRA2002}, but 
in  the one-dimensional case considered in this paper the thermal  power is 
\[
P=\frac{\pi}{6} \frac{(k_B T)^2}{\hbar},
\]
so that the viscous force as given by (\ref{thermaldissipativeforce}) is twice the 
Doppler force~\cite{JaekelReynaudPLA93}. As discussed in the next section, a similar result holds 
for coherent states with Dirichlet BC.

\section{Coherent state}

The coherent state of amplitude $\alpha$ is defined as 
an eigenstate of the annihilation operator:
\begin{equation}\label{coherent_def}
a_{\omega}\,|\alpha\rangle = \alpha\,\delta(\omega-\omega_0)|\alpha\rangle,
\end{equation}
where $\omega_0>0$ represents the frequency  of the excited mode~\cite{Dodonov_coherent}.
The symmetrical correlation function for the Dirichlet case is 
computed from (\ref{phixomega}) and (\ref{coherent_def}): 
\begin{equation}\label{coherent_correlation}
\sigma_D(\omega,\omega')=
\sigma_D^{\rm vac}(\omega,\omega')+\sigma_D^{\rm I}(\omega,\omega')+\sigma_D^{\rm II}(\omega,\omega'),
\end{equation}
with
\begin{equation}\label{sigma_vac}
\sigma_D^{\rm vac}(\omega,\omega')=4\pi\hbar |\omega|\delta(\omega+\omega'),
\end{equation}
\begin{equation}\label{coherent_correlationI}
\sigma_D^{\rm I}(\omega,\omega')=8\pi\hbar |\omega|
|\alpha|^2\Bigl[\delta(\omega-\omega_0)+
\delta(\omega+\omega_0)\Bigr]\delta(\omega+\omega'),
\end{equation}
\begin{equation}\label{coherent_correlationII}
\sigma_D^{\rm II}(\omega,\omega')=8\pi\hbar |\omega|
\Bigl[\alpha^2\delta(\omega-\omega_0)+
\alpha^*{}^2\delta(\omega+\omega_0)\Bigr]\delta(\omega-\omega').
\end{equation}
$\sigma_D^{\rm vac}$
is the contribution of vacuum fluctuations, which are contained in the coherent
state $|\alpha\rangle.$ This term 
is also present in the 
thermal case, and corresponds to the zero-temperature limit ($\overline{n}=0$) of
(\ref{sigmaD}). 
It originates from the commutator $[a_{\omega},a_{\omega'}^{\dagger}]$
as given by (\ref{commutator}). As expected,  
$\sigma_D=\sigma_D^{\rm vac}$ when $\alpha=0,$ 
since $|\alpha=0\rangle$ is the vacuum state. 
The normally-ordered correlation function is 
the sum of 
$\sigma_D^{\rm I}$ and 
$\sigma_D^{\rm II}.$ The former 
originates from 
terms of the form $a_{\omega}^{\dagger} a_{\omega'},$
whereas the latter originates from terms 
$a_{\omega} a_{\omega'}$ and their hermitian conjugates.

We calculate the force for Dirichlet BC from 
(\ref{Fomega2}) and (\ref{sigma_vac})--(\ref{coherent_correlationII}),
and then for Neumann BC from 
(\ref{forcacorrelacoes}) and (\ref{relation}). 
We write separately the contributions  from each term in 
(\ref{coherent_correlation}):
\begin{equation}\label{coherent_forca}
{\cal F}_{\stackrel{D}{\scriptscriptstyle N}}(\omega)=\chi_{\rm vac}(\omega)\delta Q(\omega)
+{\cal F}^{\rm I}(\omega)\pm{\cal F}^{\rm II}(\omega),
\end{equation}
with $\chi_{\rm vac}(\omega)$ given by (\ref{creg}) and
\begin{equation}\label{DeltaF1}
{\cal F}^{\rm I}(\omega)=\frac{4 i \hbar \omega_0}{\pi} |\alpha|^2\omega \delta Q(\omega),
\end{equation}
\begin{equation}\label{DeltaF2}
{\cal F}^{\rm II}(\omega)=\frac{4 i \hbar \omega_0}{\pi}\Bigl[
\alpha^2\,(\omega-\omega_0) \delta Q(\omega-2\omega_0)
+\alpha^*{}^2\,(\omega+\omega_0)\delta Q(\omega+2\omega_0)\Bigr].
\end{equation}
As expected, the contribution from $\sigma_D^{\rm vac}$ leads to the vacuum force 
already discussed in connection with the zero-temperature limit of the thermal field [see
(\ref{chivacres})]. The contribution from  $\sigma_D^{\rm I}$ is also of the form 
${\cal F}^{\rm I}(\omega)=\chi^{\rm I}(\omega) \delta Q(\omega)$ and the same for
the two types of BC, because $\sigma_D^{\rm I}(\omega,\omega')\propto \delta(\omega+\omega').$
However, 
the coherent state contains a phase reference, namely $\delta$ in 
$\alpha=|\alpha|e^{i\delta},$ which breaks the invariance under time translations.
As a consequence, the forces for Dirichlet and Neumann BC are different due to the presence of the 
term ${\cal F}^{\rm II}$ in (\ref{coherent_forca}). 
As expected, the symmetry is restored if we average over $\delta,$ i.e. if we 
replace the coherent state by the incoherent statistical mixture defined by the 
density matrix operator
\begin{equation}\label{rho}
\rho = \int_0^{2\pi} \frac{d \delta}{2\pi} ||\alpha|e^{i\delta}\rangle 
\langle |\alpha|e^{i\delta}|.
\end{equation}
In fact,  
${\cal F}^{\rm II}(\omega)$ vanishes 
if we take such an
average, as it can be checked in (\ref{DeltaF2}) or (perhaps more easily)
from the expression in the time domain:
\begin{equation}\label{DeltaF2t}
F^{\rm II}(t)=-\frac{4 \hbar \omega_0}{\pi} |\alpha|^2\Bigl[\cos(2\omega_0t-2\delta)
\delta {\dot q}(t)+\omega_0\sin(2\omega_0t-2\delta)\delta q(t)\Bigr].
\end{equation}
Since ${\cal F}^{\rm I}(\omega)$ does not depend on the phase $\delta,$ it does not
change when the phase average is taken, and reads in the time domain
\begin{equation}\label{DeltaF1t}
F^{\rm I}(t)=-\frac{4 \hbar \omega_0}{\pi} |\alpha|^2\delta {\dot q}(t).
\end{equation}

It is useful to 
calculate the power $P$ incident on the plate, in order to 
compare with the Doppler force as given by (\ref{Doppler}). 
For this end, we  calculate the average of the energy-momentum tensor 
component $T^{01}_i=\{\partial_x \phi_i(x,t),\partial_t \phi_i(x,t)\}/2$ 
considering the incident field $\phi_i(x,t)$ alone, and taken in 
normal order so as to discard the contribution of vacuum fluctuations:
\begin{equation}\label{T01p}
\langle : T^{01}_{i+} : \rangle = -\frac{\hbar \omega_0}
{2\pi} \Bigl[|\alpha|^2+{\rm Re}(\alpha^2 e^{-2 i \omega_0 (t+x)})\Bigr]
\end{equation}
for $x>0,$ and 
\begin{equation}\label{T01m}
\langle : T^{01}_{i-} : \rangle = \frac{\hbar \omega_0}{2\pi} 
\Bigl[|\alpha|^2+{\rm Re}(\alpha^2 e^{-2 i \omega_0 (t-x)})\Bigr]
\end{equation}
for $x<0.$ 
The total incident power is 
\begin{equation}\label{P_coh}
P = \langle : T^{01}_{i-} : \rangle(x=0^-) - \langle : T^{01}_{i+} : \rangle(x=0^+)=
\frac{\hbar \omega_0}
{\pi}|\alpha|^2 \left[1+\cos\left(2\omega_0t-2\delta\right)\right].
\end{equation}
We find the Doppler force for a coherent state by replacing (\ref{P_coh}) 
into (\ref{Doppler}). For Dirichlet BC, the sum of the viscous terms
 in (\ref{DeltaF2t}) and (\ref{DeltaF1t}) is twice the Doppler force, as in the case 
of the thermal field for both BC. This  also holds  for Neumann BC only if we average over the 
phase $\delta.$ 

For a pure coherent state the force
does not vanish when the plate is at rest, due to the
second term in (\ref{DeltaF2t}), which is
 proportional to the instantaneous position of the plate.
This is a consequence of the spatial dependence of the energy-momentum tensor of
the incident field [see (\ref{T01p}) and (\ref{T01m})].  On the other hand, for the thermal state and
the incoherent mixture given by (\ref{rho}), translational symmetry forbids the existence of a net average 
force on a single static mirror. 

\section{Conclusion}

Dirichlet and Neumann BC yield the same force on a moving mirror (in the nonrelativistic approximation)
when the field state is symmetrical under time translations. This is the case for  vacuum 
and thermal states, but not for a coherent state. The coherent state introduces a phase reference which breaks this
symmetry. When averaging over the phase of the coherent state, we recover identical results 
for the two BC.  

For both thermal and coherent states, the force may be split into two contributions: the first 
accounting for vacuum fluctuations,  the second representing the contribution of the normally-ordered
correlation function. The latter vanishes in the limits of zero temperature (for the thermal state) and 
of zero amplitude (for the coherent state). Thus, 
the force for the vacuum state may be recovered from these two examples as a limiting case.

In both thermal and coherent cases the force contains a dissipative component proportional to 
the velocity of the mirror (viscous force). It is tempting to interpret this effect 
as a consequence of Doppler shifting the frequencies of the incident photons. 
However, the change of amplitude and phase by reflection is also determinant.
In particular, the phase acquired by reflection underlies the difference between 
the results for Dirichlet and Neumann BC. 
No explicit connection between the viscous force and the Doppler effect holds for 
the coherent state with Neumann BC, whereas in the other examples discussed in this paper  
the viscous force was found to be twice the Doppler force.  

C.F. and P.A.M.N. thank CNPq for partial financial 
support. P.A.M.N.
also acknowledges support by
PRONEX and the Brazilian Institute for Quantum Information.

\vskip 2.0 cm
\centerline{\bf APPENDIX: Correlation function for a thermal state}
\bigskip
In this Appendix we compute the correlation function
$\sigma_D(\omega,\omega')$ for a thermal field, 
used in Section 3 for deriving the force.
From eq.~(\ref{phixomega}) we find 
\begin{equation}
\sigma_D(\omega,\omega')=C(\omega,\omega')+C(\omega',\omega)
\end{equation}
with
\begin{equation}
C(\omega,\omega')=
4\pi\hbar\,\sqrt{\omega\vert\,\omega'\vert}\;
\theta(\omega)\theta(-\omega')
\langle a_{\omega}\, a^\dagger_{-\omega'}\rangle\;+\; 
4\pi\hbar\,\sqrt{\vert\omega\vert\,\omega'}\;
\theta(-\omega)\theta(\omega')
\langle a^\dagger_{-\omega}\, a_{\omega'}\rangle.
\end{equation}
Using the thermal correlations
\begin{eqnarray}
\langle a_{\omega}\, a^\dagger_{-\omega'}\rangle&=&\nonumber
\biggl(\overline n(\omega)+1\biggr)\delta(\omega+\omega')\cr\cr
\langle a^\dagger_{-\omega}\, a_{\omega'}\rangle&=&\nonumber
\;\overline n(\omega')\;\delta(\omega+\omega'),
\end{eqnarray}
where $\overline n(\omega)$ is the average thermal photon number defined in 
(\ref{nT}), 
we find
\begin{equation}\label{sigma_app}
\sigma_D(\omega,\omega')=4\pi\hbar\left[\omega\theta(\omega)
\biggl(1+2\overline n(\omega)\biggr)+
\omega'\theta(\omega')
\biggl(1+2\overline n(\omega')\biggr)\right]\,\delta(\omega+\omega').
\end{equation}
Finally, we use that $n(-\omega)=n(\omega)$ and 
$\omega(\theta(\omega)-\theta(-\omega))=|\omega|$ to derive 
(\ref{sigmaD}) from (\ref{sigma_app}).


\begin{thebibliography}{99}

\bibitem{Casimir}H. B. G. Casimir, Proc. K. Ned. Akad.\ Wet. \textbf{51},
793\textbf{ }(1948).

\bibitem{Boyer} T.H. Boyer, Phys. Rev A {\bf 9}, 2078 (1974); V. Hushwater, Am. J. Phys. 
{\bf 65}, 381 (1997); M.V. Cougo-Pinto, C. Farina, F. Santos and A. Tort, J. Phys. A {\bf 32}, 4463 (1999); 
M.V. Cougo-Pinto, C. Farina and A. Ten\'orio, Braz. J. Phys. {\bf 29}, 371 (1999).


\bibitem{permeableplatesrecently} O. Kenneth, I. Klich and M. Revzen, Phys. Rev. D 
{\bf 65}, 045005 (2002); O. Kenneth, I. Klich, A. Mann  and M. Revzen, Phys. Rev. Lett. {\bf 89}, 033001 (2002); D. Iannuzzi and F. Capasso, Phys. Rev. Lett. {\bf 91}, 029101 (2003); 
 O. Kenneth, I. Klich, A. Mann  and M. Revzen, Phys. Rev. Lett. {\bf 91}, 029101 (2003); 
M.V. Cougo-Pinto, C. Farina, F. Santos and A. Tort, Phys. Lett. B {\bf 446}, 170 (1999); D.T. Alves, C. Farina and A.C. Tort, Phys. Rev. A {\bf 61}, 034102 (2000); 
 D.T. Alves, F.A. Barone, C. Farina and A.C. Tort,  Phys. Rev. A {\bf 67},  022103 (2003).

\bibitem{Boyer1+1} J. H. Cooke, Am. J. Phys. {\bf 66}, 569 (1998).

\bibitem{reviewstat} G. Plunien, B. Muller and W. Greiner, Phys. Rep. {\bf 134}, 89 (1986); 
 V.M. Mostepanenko and N.N. Trunov, {\it The 
Casimir Effect and Its Applications}, (Clarendon Press, Oxford, 1997); 
 M. Bordag, U. Mohideen and V.M. Mostepanenko, Phys. Rep. {\bf 353}, 205 (2001); P. W. Milonni, {\it The Quantum Vacuum}, (Academic Press, Boston, 1993).

\bibitem{Moore}G. T. Moore, J. Math. Phys. \textbf{11}, 2679 (1970).

\bibitem{dissipation} V. B. Braginsky and F. Ya. Khalili,
Phys. Lett. A {\bf 161}, 197 (1991); 
M. T. Jaekel and S. Reynaud, Quantum Opt. {\bf 4}, 39 (1992); 
G. Barton, {\it New aspects of the Casimir effect: fluctuations and 
radiative reaction,} in {\it Cavity Quantum Electrodynamics,} Supplement: Advances in 
Atomic, Molecular and Optical Physics, edited by
P. Berman (Academic Press, New York, 1993);
P. A. Maia Neto and S. Reynaud, Phys. Rev. A {\bf 47}, 1639 (1993).

\bibitem{emission}  One plate: G. Barton and C. Eberlein, 
Ann. Phys. (N.Y.) {\bf 227}, 222 (1993);
 P. A. Maia Neto and L. A. S. Machado, Phys. Rev. A \textbf{54,
}3420 (1996);
R. G\"{u}tig and C. Eberlein, J. Phys. (London) A: Math. 
Gen. {\bf 31}, 6819 (1998);
J. P. F. Mendon\c ca, P. A. Maia Neto and F. I. Takakura, Optics Communications {\bf 160}, 335 (1999).
Two plates or cavity: A. Lambrecht, M.-T. Jaekel and S. Reynaud, Phys. 
Rev. Lett. {\bf 77}, 615 (1996); V. V. Dodonov, J. Phys. A {\bf 31}, 
9835 (1998); D. F. Mundarain and P. A. Maia Neto, 
Phys. Rev. A {\bf 57}, 1379 (1998);
M. Crocce, D. A. R. Dalvit and F. D. Mazzitelli, Phys. Rev. A {\bf 66}, 033811 (2002).

\bibitem{FDavies} S. A. Fulling and P. C. W. Davies, Proc. R. Soc. London \textbf{A348}, 393 (1976).
%
\bibitem{Ford} L. H. Ford and A. Vilenkin, Phys. Rev. D \textbf{25}, 2569
(1982).
%
\bibitem{PA94} P. A. Maia Neto, J. Phys. A \textbf{27}, 2167 (1994).

\bibitem{Unruh-Davies} P. C. W. Davies, J. Phys. A {\bf 8}, 609 (1975);
W. G. Unruh, Phys. Rev. D {\bf 14}, 870 (1976).

\bibitem{black_hole} S. Hawking, Commun. Math. Phys. {\bf 43}, 199 (1975). 

\bibitem{sono} J. Schwinger, Proc. Nat. Acad. Sci. USA \textbf{90}, 958 (1993); C. Eberlein,
Phys. Rev. Lett. \textbf{76}, 3842 (1996).

\bibitem{mass} One plate: M.-T. Jaekel and 
S. Reynaud, Phys. Lett. A {\bf 180}, 9 (1993);
G. Barton and A. Calogeracos, Ann. Phys. (NY) {\bf 238},
227 (1995); A. Calogeracos and 
G. Barton, {\it ibid.} {\bf 238}, 268
(1995). Two plates: 
M.-T. Jaekel and S. Reynaud, J. Phys. I {\bf 3}, 1093 (1993);
L. A. S. Machado and P. A. Maia Neto, Phys. Rev. D {\bf 65}, 125005 (2002).


\bibitem{PAprl} D. Dalvit and P. A. Maia Neto, Phys. Rev. Lett. \textbf{84}, 798 (2000); 
 P. A. Maia Neto and D. Dalvit, Phys. Rev. A \textbf{62}, 042103 (2000).

\bibitem{review} M.-T. Jaekel and S. Reynaud, Rept. Prog. Phys. {\bf 60}, 863 (1997); 
R. Golestanian and M. Kardar, Rev. Mod. Phys. \textbf{71,
}1233 (1999).

\bibitem{JaekelReynaudPLA93} M. T. Jaekel and S. Reynaud, Phys. Lett. A {\bf 172}, 319 (1993).
%
\bibitem{PRA2002} L. A. S. Machado, P. A. Maia Neto and C. Farina, Phys. Rev. D {\bf 66}, 105016 (2002). 

\bibitem{Dodonov_coherent} The dynamics of a coherent state of the field inside a cavity with a moving mirror was
considered by V. V. Dodonov, A. Klimov and V. I. Man'ko [Phys. Letters A {\bf 149} 225 (1990)]. For a single moving mirror, 
the modification of the state of the field may be neglected when computing the force. Hence we take 
a prescribed field state. 

\end{thebibliography}
\end{document}